\documentclass{elsart}
\usepackage{epsfig}

\usepackage{amssymb}

\begin{document}

\begin{frontmatter}
\title{Sources of Stellar Energy, Einstein- Eddington Timescale of Gravitational Contraction and
Eternally Collapsing Objects}

\author[label1]{Abhas Mitra}
%\author{Abhas Mitra}
\address[label1]{Max Planck Institut fur Kernphysik, Saupfercheckweg 1, D- 69117 Heidelberg, Germany}
\address[label2] {also, Theoretical Astrophysics Section, BARC, Mumbai
-400085, India}
%\begin{document}

\renewcommand\b{\begin{equation}}
\renewcommand\e{\end{equation}}
\renewcommand\l{\left}
\renewcommand\r{\right}

\begin {abstract} We point out that although conventional stars are
primarily fed by burning of nuclear fuel at their cores, in a strict
sense, the process of release of stored gravitational energy, known
as, Kelvin - Helmholtz (KH) process  is either also operational
albeit at an arbitrary slow rate, or lying in wait to take over at
the disruption of the nuclear channel.
   In fact, the latter mode of energy release is the true feature of
 any self-gravity bound object including stars. We also highlight
the almost forgotten fact that Eddington was the first physicist to
introduce Special Relativity into the problem and correctly insist
that, actually, total energy stored in a star is not the mere
Newtonian energy but the total mass energy ($E = M c^2$).
Accordingly, Eddington defined an ``Einstein Time Scale'' of
Evolution where the maximum age of the Sun turned out to be $t_E
\approx 1.4 \times 10^{13}$ yr. This concept has a fundamental
importance though we know now that Sun {\em in its present form}
cannot survive for more than 10 billion years.
   We
extend this concept by introducing General Relativity
   and show that
the minimum value of depletion of total mass-energy is $t_E =\infty$
not only for Sun but  for and sufficiently massive or dense object.
We propose that this time scale be known in the name of ``Einstein -
Eddington''.  We also point out that, recently, it has been shown
that as massive stars undergo continued collapse to become a Black
Hole, first they become extremely relativistic Radiation Pressure
Supported Stars. And the life time of such relativistic radiation
pressure supported compact stars is indeed dictated by this Einstein
-Eddington time scale whose concept is formally developed here.
Since this observed  time scale of this radiation pressure supported
quasistatic state turns out to be infinite, such objects are called
 Eternally Collapsing Objects (MECO). Further since ECOs are
 expected to have strong intrinsic magnetic field, they are also
 known as ``Magnetospheric ECO'' or MECO.
\end{abstract}

\end{frontmatter}

\section{Introduction}

A star is a self-luminous self-gravitating object which is evolving
all the time,  howsoever slow the rate may be. Though it is in
quasi-static hydrodynamical equilibrium, because of its constant
thermodynamical evolution, in the {\em strictest sense}, $d R_0/d t=
{\dot R}_0 \neq 0$, where $R_0$ is its radius. If the adiabatic
index of the stellar fluid is $\Gamma$, then, virial theorem yields,
the internal energy as  (Kippenhahn \& Weigert 1990, Chandrasekhar
1967)
\begin{equation}
 U = -{1\over {3(\Gamma -1)}} \Omega
\end{equation}
 where
 \begin{equation}
 \Omega = -{3\over {5
-n}} {G M^2 \over {R_0}}
 \end{equation}
  is the Newtonian gravitation potential
energy of the system. Here $M$ is the (gravitational) mass of the
system. In an approximate manner, the star here is  represented by
an polytrope of degree $n$: \begin{equation} p = K \rho^{1 +1/n}
\end{equation} The foregoing equation refers just to an assumed
uniform equation of state ($K, n = constant$) all over the star and
the index $n$ is not necessarily equal to the ratio of specific
heats $\Gamma \neq 1 +1/n$. One would have $\Gamma = 1+1/n$ only if
the system  would be assumed to evolve adiabatically.

The Newtonian total energy of the system is \begin{equation} E_N =
U+ \Omega = {3\Gamma -4\over {3(\Gamma -1)}} \Omega \end{equation}
If one assumes $\Gamma$ to remain almost constant during the slow
evolution, the rate of change of the (Newtonian energy) of the
system is \begin{equation}{d E_N \over d t} = {3\Gamma -4\over
{3(\Gamma -1)}} {d \Omega \over d t} \end{equation} From Eq.(2), we
have
\begin{equation}{d \Omega \over  dt} = {3 \over 5 -n} {G M^2 \over
{R_0}^2} {d R_0 \over dt} \end{equation} Here, implicit assumptions
are that $n$ and $M$ remain constant during this slow evolution. In
the context of Newtonian physics, the latter assumption is perfectly
justified. The luminosity of the star due to gravitational
contraction, known as Kelvin - Helmholtz process is
\begin{equation} L_{KH} = - {dE_N\over dt} = {(3\Gamma -4) \over (5
-n)  (\Gamma -1)} {G M^2 \over {R_0}^2} (-\dot {R}_0) \end{equation}
Since ${\dot R}_0 <0$ during contraction, $L_{KH} > 0$. If there
would not be additional sources of luminosity, this process also
defines a natural time scale of contraction (Kippenhahn \& Weigert
1990, Chandrasekhar 1967)
\begin{equation} t_{KH} = {E_N \over L_{KH}} = {\Omega \over
d\Omega/dt} = {R_0 \over -{\dot R}_0} \end{equation}

Having made this introductory theoretical background, we shall
specifically point out the  role of both nuclear energy generation
and KH- energy generation in Sun. Then we shall highlight the
important Special Relativistic concepts introduced into the problem
for the first time by Eddington. Following the fundamental concept
of a maximal luminosity, developed by none other than Eddington, we
shall introduce General Relativity in the problem.
\section {Nuclear Fuel Supported Time Scale}
The Eq.(7) shows that, even without burning of any nuclear fuel,
there could be stars, or in a broad sense, self-gravitating objects
of finite temperature and luminosity. This is the reason that the
massive primordial clouds have finite pressure and temperature and
do not undergo any radiationless catastrophic collapse. On the other
hand, they may keep on evolving (contracting) quasistatically for
durations much larger than free fall times. Such clouds always have
a finite luminosity though the frequency of the emitted radiation
would be far below the optical range until the final stages. After
the final stages, the central region of the cloud could become hot
enough to shine in the visible optical range to appear as ``Stars''.
Actually, these are ``Pre-main -sequence'' stars, although, before
the development of modern theory of hydrogen burning stars, one
would not distinguish between main-sequence and pre-main-sequence
stars. The rise of the core temperature of these stars generating
energy by purely gravitational process would eventually ignite  the
central nuclear fuel and give birth to normal main-sequence stars.

On the other hand, very low mass stars, called Brown Dwarfs, would
continue to shine exclusively and permanently by the KH-process
(Kumar 1962, 1963a,b). Earlier it was believed than Sun too
generated energy only by this channel, and, if so, Sun's age would
be (Chandrasekhar 1967)
 \begin{equation}
 t_{KH} \approx 1.59 \times 10^7 \times q ~{\rm yr}
 \end{equation}
 where $q = 3/(5-n)$. If further, one would assume, $n =3/2$ for Sun, one would have
 $t_{KH} = 2.4 \times 10^7$ yr. We know it too well that the actual age of Sun, $t_\odot$, is
 higher by more than two orders of magnitude and obviously there is atleast another important
 source of energy generation in Sun. This statement is often misinterpreted by
 stating that ``the Kelvin - Helmholtz'' contraction hypothesis is incorrect for
 Sun. In fairness, this hypothesis is incorrect only insofar as it denies presence
 of other energy generation modes; but the basic process of KH energy generation is
 no hypothesis and is a fundamental result of astrophysics as seen by Eqs.(1-7).
 Since,  {\em in the strictest sense}, Sun is evolving and $\dot {R}_0 <0$, the equation (7)
 is always operational though the value of actual $|\dot {R}_0| \ll |\dot {R}_0^{KH}|$ where
 $\dot {R}_0^{KH}$ would be the rate of contraction in the absence of any nuclear
 energy generation, i.e., the one indicated by Eq.(8). Note that,
 \begin{equation}
 {|{\dot R}_0| \over |{{\dot R}_0}^{KH}|}  \sim {t_{KH} \over t_{\odot}} \sim 10^{-3}
 \end{equation}
\section {Enter Eddington}
Although the theory of nuclear energy generation in stellar core
culminated through the landmark work of Bethe (1939), the very
concept that some non-gravitational ``sub-atomic'' process is
responsible for Sun's energy generation was due to Eddington
(Eddington 1920, 1926). Eddington mentioned of ``transmutation of
elements'' and was the first to recognize that such processes
involved application of $E = Mc^2$ formula. Thus he, for the first
time wrote that `` {\em stars burn mass}'' and, ``{\em any radiation
is radiation of mass}'':
\begin{equation}L = -{dE\over dt} = {- d (M c^2)\over dt} \end{equation} With
this generic and special relativistic definition of luminosity, the
maximum age of the luminous phase, i.e., the time for depletion of
entire available energy of a star is \begin{equation} t_E \sim {Mc^2
\over L} = {M\over -dM/dt} \end{equation} irrespective of the {\em
model and theory of energy generation}. As per Bowers and Deeming
(1984), this time scale is called the ``Einstein Time Scale''. For
the Sun, one can easily see that
\begin{equation}t_E \sim {M_{\odot} c^2 \over L_{\odot}} \approx 1.4
\times 10^{13} ~{\rm yr} \end{equation} The very fact that there
could be a natural time scale $t_E \gg t_{KH}$ strongly gave boost
to the theories of stellar energy models different from (purely
Newtonian) KH-model. If the ``transmutation'' process, conceived by
Eddington, were operative over the entire domain of Sun (i.e., at
any $R$) and its efficiency, accumulated over the entire (current)
life time of the star would be unity (i.e., entire mass would be
converted into energy over the entire life span), the actual value
of $t_\odot$ would have been equal to $t_E$ as long as one would not
invoke GR. The fact that, actually, $t_{\odot} \sim 10^{-3} t_E$
only means that the efficiency of the process, averaged over the
entire volume, and integrated over the current life span of Sun is
accordingly smaller, $\epsilon \sim 10^{-3}$.
\section{Termination of Thermonuclear Energy Generation}
It is widely mentioned  that once Sun would stop thermonuclear
energy generation, its internal energy $U \approx 0$ until quantum
effect would give rise to new source of internal energy at much
higher density. Consequently, it is believed that, the pressure
would immediately drop significantly, $p \approx 0$, and Sun would
undergo near {\em free fall} to collapse to a point in a time
(Kippenhahn \& Weigert 1990)
\begin{equation}\tau_c= {\pi\over 2} \left({3\over 8 G \pi
\rho(0)}\right)^{1/2}  = {\pi\over 2} \left({R^3\over 2 G M
}\right)^{1/2}
\end{equation}
 where $\rho(0)$ is the central density at $t=0$ when the star is assumed
to be at rest. Here it is assumed that the star is of uniform
density. With its present density, one would have $\tau_c \approx
28$ min for Sun. This result is clearly incorrect because
 we have
found that even without any thermonuclear energy support Sun can
evolve quasistatically for atleast $2.4 \times 10^7 {\rm yr}$
(Eq.[9])! Therefore the {\em assumption of free fall} in this case
{\em undermines} the contraction time scale by a {\em factor of
atleast} $7 \times 10^{11}$.

Interestingly, if one applies GR for this problem, and assumes free
fall, one would obtain {\em exactly} the same formula for collapse
time; however, in this case $\tau_c$ would be the proper time of
collapse recorded by a comoving observer (Misner, Thorne \& Wheeler
1973).

We will see clearly that {\em absence of thermonuclear energy
generation in any self-gravitating system does not mean absence of
pressure and internal energy}. On the other hand, absence of
non-gravitational energy generation only means that the system would
now be self-consistently dictated by pure gravity. Note that a piece
of smouldering ember may be also considered as isolated and
self-luminous. But gravity plays no role here and depending on the
mass of the ember and other chemical details, the life-time of
self-luminosity could be very small or relatively longer. Central
thermonuclear energy generation is philosophically akin to the
process of burning of the ember once we do not worry about issues
like what confines and ignites either the thermonuclear fuel in the
Sun or the ember atoms. The piece of ember ceases to shine after the
exhaustion of chemical fuel because {\em it is not supported by long
range universal attractive self-gravity with a negative specific
heat}.

On the other hand, the KH process is the true and active signature
of gravitational compression and resultant energy generation.
Philosophically, one may see the KH process as the conversion of
mass into radiation by constant self-gravitational squeezing. When a
star is supported by thermonuclear energy generation, the value of
${\dot R}_0$ may be practically considered zero compared to its
value in a purely KH-phase. And if the thermonuclear energy
generation suddenly stops, definitely, atleast, initially, the star
would try to collapse.  Note that $U$ in Eq.(1) does not, per se,
 depend on any nuclear energy generation. Moreover $U >0$ for all physical systems. Then by differentiating Eq.(1), we obtain
 \begin{equation}
 { dU\over dt} = -{1\over 3(\Gamma -1)} ~{d\Omega\over dt}
 \end{equation}

 Using Eq.(6) into  Eq.(15), we have,
\begin{equation}
 { dU\over dt} = {1\over (\Gamma -1)} ~{1 \over (5 -n)} {G M^2 \over {R_0}^2} (-{\dot R}_0)
 \end{equation}
 By using Eqs. (1) and (2) in the above one, we rewrite
 \begin{equation}
 { dU\over dt} =U ~{-{\dot R}_0\over R_0}  =U ~{|{\dot R}_0|\over R_0}>0
 \end{equation}
Thus as $|{\dot R}_0|$ would suddenly increase due to absence of
thermonuclear energy generation, there would be fresh addition of
internal energy. This clearly shows that there cannot be any strict
pressure free collapse in general and even if one would assume
 the contrary, fresh supply of internal energy may  restore
quasistatic contraction. However, there could be almost free fall
only for the idealized limiting case of $\Gamma = 4/3$ when the
increase in the value of $U$ is exactly offset by the the decrease
in the value of $\Omega$ and the system always has $E_N =0$. But a
finite system having finite particle momenta always has $\Gamma
>4/3$ (Mitra 2006a). Yet there could be situations when one may indeed
have $\Gamma \approx 4/3$ for a certain duration.

 Thus self-gravitation driven KH process causes
a negative feed back by creating more pressure and internal energy.
In other words, {\em squeezing by self-gravity creates its own
antidote and prevents a runaway squeezing} in the long run. Hence,
if central thermonuclear reaction would stop in Sun, sooner or
later, it would enter into a self-gravity driven KH-mode {\it a la},
a premain -sequence-star  or any other self-gravitating object.
However, the actual luminosity of Sun, in this phase, need not be at
all equal $L_{\odot}$. On the other hand the actual value of $L$
would be self-consistently determined by the solution of coupled
collapse equations in tandem with thermodynamics and other physical
laws.

In fact, recently it has been shown that there is an approximate
generic relationship between radiation and rest mass energy density
density of any self-luminous self-gravitating object: $\rho_r/\rho_0
\approx \alpha ~z$, where $z$ is the surface gravitational redshift,
$\rho_0$ is the rest mass energy density and $\alpha =L/L_{ed}$.
Here $L$ is the luminosity of the object and $L_{ed}$ is the
corresponding Eddington luminosity (Mitra 2006a). Further it has
been shown that as the compactness $z$ increases, $\alpha \to 1$ and
the object may be supported by self-luminosity because of the KH
process.
\subsection{Physical Mechanisms for Energy Generation}
One may wonder how a cold giant molecular cloud at an initial
temperature $\sim 10$ K may generate ``heat'' during gravitational
contraction in the absence of liberation of any nuclear or chemical
energy. This happens because of excitation/deexcitation of molecular
vibrational and rotational levels. Another way of seeing this could
be in terms of time dependent (electrostatic) inter/intra molecular
interactions. From a gross macroscopic view point, one may
understand this by recalling that there is really no ``perfect
fluid'' and heat/radiation must be generated during collapse because
of various dissipative processes. More technically, such radiative
processes  for the cold cloud are  due to ``bound-bound''
interactions.

When the cloud would get partially ionized, there would be, in
addition, ``free-bound'' radiative processes. And if the fluid would
become completely ionized, the dominant radiative process would be
``free-free'' process or Bremsstrahlung. And in case, the fluid
would become extremely compact due to self-gravity, the charged
particles (as well as non charged particles including neutrinos and
photons) would tend to move in nearly circular orbits due to
gravitational bending. In such a case, the dominant radiation
mechanisms would be {\em Relativistic Gravitational Bremsstrahlung}
(Peters 1970). Along with electromagnetic processes, both weak and
strong interactions too may play their respective roles during
Bremsstrahlung. At the same time if there would be a strong
intrinsic magnetic field in the fluid, radiation will be generated
by {\em Relativistic Gravitational Synchrotron} process (Misner et
al. 1972) or cyclotron process.

 It may be recalled that, from microphysics point of view, any
 nuclear or chemical ``burning''  too is implemented by such basic
 processes like ``free -bound'' or ``free-free''.
\section{Einstein- Eddington Time Scale}
Eddington also introduced another fundamental concept in
astrophysics: a radiating object has a maximum luminosity at which
the outward push of the radiation force on the plasma would just
counterbalance the inward pull of gravity. This luminosity, known
after his name is defined as
\begin{equation}
L_{ed}(R) = {4 \pi G M(R) c\over \eta \kappa}
\end{equation} where $\eta(R) = L(R)/L$ and $\kappa$ is the average
specific opacity for a region $R$. At the boundary of the star, by
definition $\eta =1$  and $M(R) =M$  so that
\begin{equation}
L_{ed} = {4 \pi G M c\over  \kappa}
\end{equation}

Further if we define a parameter \begin{equation} \alpha = {L\over
L_{ed}} \end{equation} we will have
\begin{equation}L  = {4 \pi G M c \alpha \over  \kappa}
\end{equation}
 In case, a Newtonian star would radiate at its maximal limit, i.e., if one would
have  $\alpha =1$, then one would obtain the {\em lowest} possible
value of \begin{equation} t_{EE}^{min} = { \kappa c \over 4\pi G }
\end{equation} Further, for propagation of photons within physical
matter, the minimum value of $\kappa$ is given by :
\begin{equation}\kappa = \kappa_T =\sigma_T/ m_p \end{equation} where
\begin{equation}\sigma_T = \left({8\pi\over 3}\right) \left({e^2\over m_e
c^2}\right)^2 \end{equation}
 is the Thomson cross-section; $m_e$ is electron rest mass, $m_p$ is the proton rest mass,
 $e$ is electron charge. The numerical value of $\kappa_T = 0.4$ g cm$^{-2}$
 and this is the {\em lowest} value of $\kappa$ (Kippenhahn \& Weigert 1990).
 On the other hand,
 typical stellar opacities could be thousand times larger than this lowest value. Therefore,
 we find that, the lowest value of
 \begin{equation}
t_{EE}^{min} = { \sigma_T  c \over 4\pi G ~m_p} \end{equation}
  Using Eq.(24) in (25), we see that
\begin{equation}t_{EE}^{min} = \left({2 c \over 3 G m_p }\right) \left({e^2\over m_e
c^2}\right)^2 \end{equation}

  It is interesting that the smallest value
of the time to deplete the  mass-energy depends only on fundamental constants.
Moreover, it is independent of the (gravitational) mass of the source of radiation.
  The corresponding  numerical value  of
\begin{equation}t_{EE}^{min}  \approx 1.5 \times 10^{16}~s \sim 5 \times 10^8
~{\rm yr} \end{equation} irrespective of the mass of the star.  This
suggests that, if one would apply Special Theory of Relativity
(STR), a self-luminous phase for self-gravitating objects must exist
for atleast $5\times 10^8$ yr.

In case the mass energy loss would be because of $\nu$-emission
instead of photon emission, the opacity would be extremely small and
one may have $t^{min}_{EE}$ as small as $10$ s. Recall that this is
indeed the observed time scale of the $\nu$-burst from SN 1987A. On
the other hand, the corresponding free fall time scale of the proto
neutron star (NS) is as small as $\approx 0.1$ ms. Thus, the free
fall assumption is lower, in this case, by a factor of $\approx
10^5$.

The time scale $t_{EE}^{min}$ actually refers to the mass energy
depletion. The NS born in the SN explosion continues to cool (i.e.,
lose mass-energy) primarily through photon emission for millions of
years. In a strict sense, the cooling time scale could easily be
what is indicated by Eq.(27) eventhough intermediate $\nu$-driven
time scale could be extremely shorter.
\section {Enter General Relativity}
The physics of Sun or anything else having associated mass energy,
in the strict classical sense, is  determined by not Special
Relativity, but by GR. The formal entry of GR counterpart of Eq.(11)
was facilitated with the discovery of the spacetime structure
exterior to a radiating star by  Vaidya (1951). At the boundary of
the star, one has
\begin{equation}
ds^2 = (1- 2G M/R_0 c^2) du^2 + 2 du\, dR - R_0^2(d\theta^2 +
\sin^2\theta d\phi^2)
\end{equation}
where  $u$ is the retarded time, $\theta$ is the polar and $\phi$ is
the azimuth angle. As the collapse/contraction proceeds both
$M(R_0)$ and $R_0$ decrease. The luminosity of the star as seen by a
{\em distant} observer is
\begin{equation}
L^\infty = -{d (M c^2)\over du}
\end{equation}
It may be reminded here that gravitational mass $M c^2$ is the total
mass-energy content seen by a distant observer and not by a local
observer. On the other hand, in GR, the spacetime around a {\em
strictly static} object is given by the radiationless {\em vacuum}
Schwarzschild metric which corresponds to both
\begin{equation}
L _{local} = L^{\infty} =0
\end{equation}
This very property should disqualify {\em strictly static} objects
to acquire the nomenclature of a ''Star'', and it would be more
reasonable to call them as ``Objects'' only. Since there is
mass-energy associated with any radiation,  by definition, {\em
strictly static} GR objects must not be self-luminous, and, one must
have temperature $T=T_s =0$ at the boundary. But if $T$ is finite
just beneath the boundary, the associated radiation field must
penetrate the boundary too (unless the internal region is a trapped
one) and hence {\em strictly static} GR objects must have $T=0$
everywhere. Hence, there could be {\em strictly static} GR objects
only for perfectly degenerate and absolutely cold material.

For relativistic objects, one important observable measure of the grip of gravity
on the surface is the surface gravitational redshift (of spectral lines emanating from the surface)

\begin{equation}
 z= (1- 2 GM/R_0 c^2)^{-1/2} -1
 \end{equation}

For this latter external metric, a {\em strictly static}
self-gravitating object has a fundamental constraint $z <2$
(Buchdahl 1959). However, when the external spacetime contains
radiation, i.e., when it is to be described by Vaidya metric, there
is no such constraint : $z \le \infty$. This infinite difference in
the limiting value of $z$ depending on whether the external
spacetime is luminous or not is indeed a very subtle point reminding
one of the epithet ``Subtle is the Lord'' in the context of GR. Let
us recall here that very massive stars or any sufficiently dense
object would undergo continued gravitational collapse to $z=\infty$
Black Hole (BH) stage.

 As discussed earlier, the minimum self-luminous time scale is obtained when
 $L= L_{ed}$. And GR causes the local value of $L_{ed}$ to increase from its Newtonian value by a
 factor of $(1+z)$ (Mitra 1998a, Mitra 2006a):
\begin{equation}
L_{ed} = {4 \pi G M c\over \kappa} (1+z)
\end{equation}
However, because of  the joint effect of gravitational redshift and
gravitational time dilation, the distant observer sees a reduced
Eddington luminosity: \begin{equation}L_{ed}^\infty = {L_{ed} \over
(1+z)^2} = {4 \pi G M c\over \kappa (1+z)}
\end{equation}
Consequently, the minimum value of previously defined Einstein-
Eddington Time Scale, in GR, becomes \begin{equation} u_{EE}^{min} =
{ Mc^2 \over L_{ed}^\infty} = {k c \over 4 \pi G} (1+z)
\end{equation} Since in principle, during continued collapse, $z \to
\infty$, clearly, the Einstein- Eddington time scale for depletion
of mass energy becomes infinite for {\em arbitrary} value of the
opacity $\kappa$:
\begin{equation}u_{EH}^{min} =\infty \end{equation} This immediately shows that
{\em irrespective of the actual value} of mass of the
self-gravitating object, the phenomenon of self-luminosity becomes
eternal.

 Considering the minimum value of photon opacity $\kappa = \sigma_T/ m_p$ in  equation (34), we have
\begin{equation}u_{EE}^{min} = {\sigma_T c\over 4\pi G m_p} (1+z) \end{equation} In terms of
fundamental constants, we eventually obtain
\begin{equation}u_{EE}^{min}  =\left({2 c \over 3 G m_p
}\right) \left({e^2\over m_e c^2}\right)^2 (1+z) \approx 5 \times
10^8 ~ (1+z) ~{\rm yr} \end{equation}

As discussed in the previous section if there would be a GR collapse
and formation of a compact object, it is likely that the opacity
would be determined by $\nu$-emission. In such a case, in Eq.(34),
one should use appropriate $\nu$ opacities rather than $\sigma_T$.
The resultant time scale in such a case would be $u_{EE}^{min} > 10
(1+z)$ s. However since it is much easier to maintain a photon
mediated Eddington luminosity $\sim 10^{38}$ erg/s  instead of a
$\nu$-Eddington luminosity of atleast $10^{54}$ erg/s (for $ 1
M_\odot$ object) and the system would like to spend minimum energy,
it is likely that the eventual long terms energy depletion time
scale would be still governed by Eq.(37). Note again that, in any
case, for BH formation ($z \to \infty$), $u_{EE}^{min} \to \infty$
{\em irrespective} of the value of opacity $\kappa$.
\section{Physical Implications and Applications}
There have been some recent developments in the study of GR
gravitational collapse which show that GR collapse time scale is
indeed determined by the Einstein-Eddington time scale developed
above. Note that the massive objects tend to collapse inexorably to
BHs having an Event Horizon (EH) with $z=\infty$. If so, the object
must pass through states having {\em arbitrarily large} but finite
$z$ states to reach the $z=\infty$ state. It has been shown that as
the object would become more and more compact (i.e., $z$ would
increase), the object would become radiation energy dominated,
$\rho_r \gg \rho_0$ (Mitra 2006a). This happens because collapse
generated radiation quanta (i) Spend more time within the body
because of intense matter-radiation interaction (diffusion) and also
because (ii) they get trapped within the body because of the
extremely strong self-gravity. The density of trapped radiation
increases as $\rho_r \sim R^{-3} (1+z)^2$ (Mitra \& Glendenning
2006). The corresponding heat flux grows in a similar fashion
\begin{equation}q_{trap} \sim R^{-3} (1+z)^2
\end{equation} The GR local Eddington luminosity (Eq.[32])
corresponds to a critical outward heat flux  of
\begin{equation}q_{ed} = {L_{ed} \over 4\pi R^2} = {GMc \over
\kappa R^2} (1+z) \end{equation}

The ratio of the actual heat flux to the critical Eddington flux
grows as \begin{equation}\alpha = {q_{trap} \over q_{ed}} \sim
{(1+z) \over R M} \end{equation} Initially, of course, the value of
$\alpha \ll 1$. But as $z \to \infty$ during the BH formation, the
value of $\alpha$ starts increasing dramatically $\sim (1+z)$.
Sooner or later, at an appropriate range of finite value of $z$, one
must attain a state $\alpha \approx 1$ when the outward radiation
flux would attain its critical ``Eddington value''. By the very
definition of an ``Eddington Luminosity'', catastrophic collapse
would then degenerate into a secular quasistatic contraction
supported by radiation pressure (Mitra \& Glendenning 2006).

Even in Newtonian gravitation, there could be stars supported
entirely by radiation pressure. However, in Newtonian gravitation,
such Radiation  Pressure Supported Stats (RPSSs) turn out be
extremely massive. Recently, it has been explained that this
requirement of an excessive large mass for Newtonian RPSSs stems
from the fact that they have weak gravity ($z \ll 1$) and the only
way radiation pressure can still dominate would be if $M > 7200
M_\odot$ (Mitra 2006b) . On the other hand, if there would be
objects with appropriately large $z \gg 1$, one will have an RPSS at
arbitrary low or high value of $M$ because in such a case, $\rho_r
>> \rho_0$ as $z \gg 1$ (Mitra 2006b). In otherwords, during
continued collapse, the collapsing object first becomes an extremely
relativistic ($z \gg 1$) RPSS before becoming a true BH with
($z=\infty$). Then as explained by Eq.(37), as the RPSS tends to
become a true BH with $z \to \infty$, the lifetime of the PRSS phase
becomes infinite:
\begin{equation}u_{EE}^{min} = 5 \times 10^8~(1+z) ~{\rm
yr} \to \infty \end{equation}
 Hence such objects
have been termed as ``Eternally Collapsing Objects'' (ECO)  (Mitra
1998b,  Mitra 2000, Mitra 2002a,b,  Mitra 2006b,c,d,e,f,g,h,  Mitra
\& Glendenning 2006).  Any  astrophysical plasma is always endowed
with microscopic currents and some intrinsic magnetic field ($B$).
When such a plasma contracts, in a crude picture, its magnetic field
gets compressed as $B \sim R_0^{-2}$. And this is the basic reason
that a compact Neutron Star has strong intrinsic magnetic field.
Based on this simple argument, it was postulated by the present
author that the compact and non-singular ECOs thus must possess
strong intrinsic magnetic field (Mitra 1998b  Mitra 2000, Mitra
2002a,b). It was thus natural that a spinning ECO would have a
magnetosphere like a pulsar and accordingly, the phrase
Magnetospheric ECO or MECO  first coinded by  Leiter, Mitra \&
Robertson (2001). Subsequently Mitra withdrew his name from
revisions of this preprint which attempted to make a particular
version of MECO with {\em assumed equipartition magnetic field}
(Leiter \& Robertson 2003).

Thus the Einstein -Eddington time scale developed here is realized
in the arena of GR.
\subsection{Difference with Traditional BH Picture}
 The only exact analytical
model of BH formation involves the collapse of a homogeneous
spherical ball of dust (Oppenheimer \& Snyder 1939, OS). It is known
that, in this case, the time of formation of the Event Horizon
($z=\infty$), as seen by a distant observer, is $t_{EH} = \infty$.
On the other hand, the comoving proper time of collapse (upto the
central singularity) is still given by the Newtonian expression
(14). One may initially think that Eq.(37) exactly corresponds to
the ideal Oppenheimer-Snyder collapse yielding $t_{EH} =\infty$. But
we discuss below that such an impression is far from correct.

 To see this, in a first approximation, we may transform, $u_{EE}^{min}$ in Eq.(37) into local proper time scale which
 would be lower by a factor of $(1+z)$ because of GR time dilatation:
 \begin{equation}
 \tau_{EE}^{min} ={\sigma_T c\over 4\pi G m_p} \approx 5 \times 10^8 ~ {\rm
 yr}
 \end{equation}
In contrast, for the dust collapse, the proper comoving time scale
is essentially the  free fall time scale
\begin{equation}\tau_{ff} \sim (G \rho)^{-1/2} \end{equation} which has close similarity with
Eq.(14). Now suppose we are considering the continued collapse of a
$10 M_\odot$ mass star into a BH. During pressureless freefall,
$\rho$ would keep on  increasing as $R^{-3}$ and at one stage
density would attain nuclear value $\rho \sim 10^{14}$ g cm$^{-3}$.
At this stage one would have $\tau_{ff} \sim 10^{-4}$ s. In such a
case, we find that
\begin{equation}{\tau_{EE}^{min} \over \tau_{ff}} \sim  10^{20} ~\rho_{14}^{1/2}
\end{equation} where $\rho_{14}$ is the density in units of $10^{14}$ g cm$^{-3}$.

$\bullet$ Therefore, clearly, even though, formally, both, $u_{EH} =
t_{EH} =\infty$, there are {\em extreme latent physical differences
between the two cases}; the actual duration of the corresponding
proper time scales in the two cases could {\sl differ by a factor of
$10^{20}$ or even more} (actually $\infty$). We recall here that
even when we were considering a purely Newtonian case, the free fall
assumption {\em undermined actual} contraction time
 scale by a factor of atleast $7\times 10^{11}$ for the case of
 would be
 collapse of Sun.

 Further, suppose, we are considering the actual observational aspect of
 the contracting star at the late stages of contraction, i.e., $z \gg 1$.
 To be more specific, suppose we are considering a stage with $z =10^8$.
 Recall that, in comparison, for a Neutron Star, we have $z = 0.1 -0.2$.
 The minimum time scale of this stage would be
 \begin{equation}
 u_{EE}^{min} \approx 1.5 \times 10^{24} {\rm s} \sim 3\times 10^{6} ~ Hubble ~Time
 \end{equation}
 In contrast, for Free-Fall, one may also tentatively write the observed time scale  as
 \begin{equation}
 t_{ff} \sim (1+z) \tau_{ff}
 \end{equation}
so that, in this case, we have,
 \begin{equation}
 t_{ff}  \sim (1+z) 10^{-5}  \sim 10^3 ~{\rm s}
 \end{equation}
 because, in this case, the density would be $\rho \sim 10^{16}$ g cm$^{-3}$. However, the
 free-fall time scale, in a strict sense, is actually devoid of any physical meaning
 because a {\em dust collapse is radiationless}, and the distant observer,
 in a strict sense, would never see the collapse in the {\em absence}
 of any emitted radiation:
 \begin{equation}
 L_{ff} = L_{ff}^\infty \equiv 0
 \end{equation}
 $\bullet$ In contrast even at this  extreme late stage of radiative collapse ($z = 10^8$), a distant
 astronomer would measure a luminosity of
 \begin{equation}
 L_{ed}^\infty = 1.3 \times 10^{31} ~{\rm erg/s}
 \end{equation}
 In principle, {\em this luminosity is measurable} for a duration of $\sim 3\times 10^6$ Hubble Time
 (see Eq.[36]). In fact, the {\em quiescent luminosity
 of stellar mass BH candidates could indeed lie in this range}.
 However because of extreme redshift of MECO surface and its
 photosphere, the observed radiation would certainly not be in the
 X-ray band. On the other hand, it could be in infrared, millimeter or microwave
 band. Nonetheless, the Robertson \& Leiter  model of MECO (2006a) has some specific prediction
 on this based on several assumptions and simplifications.

 Despite such extreme physical differences between the scenarios involving
 idealized free fall and realistic radiative cases, it is often assumed that, the real
 collapsing stars may obey the rules of pressureless dust collapse. If so, then
 the luminosity of the star at late stages fall off exponentially in a fashion
 (Misner, Thorne \& Wheeler 1972)
 \begin{equation}
 L_{ff}^\infty = L_0~ \exp{\left[- {2\over 3\sqrt 3} {c~ t_{ff}\over
 R_g}\right]}
 \end{equation}
 where $R_g = 2 GM/c^2 = 3\times 10^6$cm is the  gravitational radius of the star and is
 constant in the pressureless/radiationless case. Let the value of initial luminosity of the star be $L_0 \sim 10 L_{\odot} \sim 4 \times 10^{34}$ erg/s. For the $10 M_\odot$ mass
 star,  the time constant in the exponential function is
 \begin{equation}
  {3 \sqrt {3} ~R_g\over 2c} \sim 10^{-4} ~s
 \end{equation}

 Then combining Eqs. (43), (50), and (51), it follows that, for the
 given case, we will have
 \begin{equation}
 L_{ff}^\infty \sim L_0 \exp{[ -(1+z)]} \sim L_0 \exp [-10^8]
 \end{equation}

$\bullet$ It is very clear then that even when the star will have $z
\sim 10$, let alone, $z =10^{8}$, for all practical purpose the {\em
observed luminosity} $L_{ff}^\infty \approx 0$.
 Further, for $z \gg 1$, the luminosity would be infinetisimal $L^\infty \to 0$.
 Let us clarify that here we took $z=10^8$ only as a likely case of
extreme large value of $z$. In no way do we insist that the actual
value of $z$ will be precisely this. Just like the local temperature
 $T$ of a MECO depends uniquely on $M$  (Mitra, 2006b, Mitra \& Glendenning
 2006):
 \begin{equation}
 T \approx 600 {\rm MeV} ~ \left({M \over M_\odot}\right)^{-1/2}
 \end{equation}
it is possible that $z$ too depends uniquely on $M$. As the ECO
keeps on radiating indefinitely, $M$ keeps on decreasing
indefinitely and $z$ must be evolving (increasing) all the time to
attain the $z=\infty$ BH stage. The above qualitative discussion
would remain valid for any large value of $z \gg 1$. Thus in the
present paper, neither do we really assume any specific value of $z$
nor do we purport to justify any specific $z-M$ relationship. On the
other hand, this could be the topic of a future study on the
fundamental property of ECOs.

  Even if the value
 of $L_0$ would be extremely larger than what has been considered here, the
 foregoing conclusion would remain unchanged
 for $z \gg 1$. Thus while the assumption of free fall collapse leads a scenario
 where the luminosity of the contracting  star practically becomes zero
  in a few $\tau_{ff}$, for the radiative collapse, it is possible
 that the star remains observable for practically infinite time scale eventhough
 ultimately it would eventually (mathematically) become a BH with $L^\infty =0$. In other words,
 while the effective duration of the luminous phase of a contracting star, in the
 free fall paradigm, is only few free fall time scales and thus insignificant, for
 radiative collapse the same could easily be more than the age of the Universe.

 Thus the assumption of free fall clearly leads to an incorrect picture
 for physical radiative collapse even though one obtains $u_{EH} = t_{EH}
 =\infty$ in either case. Again we recall that even when we do not consider any
 relativity, special or general, and would naively consider the present internal
 energy of the Sun as its ultimate energy source, the free fall assumption undermines
 the duration of the active luminous phase by a factor of atleast $7 \times 10^{11}$.

\section{Comparison With Other Works}
Since $u_{EE}$ refers to a {\em depletion} of mass energy, one
expects $M=0$ for the end product. Thus the BH formed asymptotically
should have a unique mass $M=0$. And it has indeed been shown that
the integration constant which appeared in the so-called vacuum
Schwarzschild solution has the unique value of zero: $\alpha_0 = 2 M
=0$ (Mitra 2005, 2006b,c,d,e). Physically, as the continued collapse
process becomes eternal, all available mass energy is radiated away.
However, a BH with $M =0$ does not necessarily mean absence of
matter or violation of any baryon/lepton number. All it means is
that total {\em positive} mass energy associated with baryons,
leptons and radiation gets exactly offset by {\em negative}
self-gravitational energy. We may see now that this profound
conclusion is indeed consistent with the phenomenon of GR collapse.

A vacuum Schwarzschild solution, actually Hilbert solution (Mitra
2006f), with a supposed $M >0$ yields the traditional BH paradigm.
On the other hand, a solution with supposed $M <0$ is equally valid
from pure mathematical perspective and would yield a ``Naked
Singularity'' (NS) without an Event Horizon (EH). Thus, a $M=0$ BH
is a borderline case between the two and  it is no spacetime
singularity at all because  it requires infinite comoving proper
time to form : $\tau =\infty$.
\subsection{Physical Interpretation of Pressureless Collapse}
First, note that, for the assumed pressureless case, {\em one
obtains
 exactly the same value} of collapse (proper) time scale for both the Newtonian
and the GR case (the same Eq.(14) remains valid in either case). But
from a strict  GR viewpoint, a purely Newtonian result is valid only
in a completely flat spacetime, i.e., when gravitation is infinitely
weak. In other words, atleast for isolated objects, GR would
approach the exact Newtonian limit
   if the mass-energy density $\rho c^2 \to
0$ everywhere. Thus we get a hint that a strict denial of pressure
probably is tantamount to denial of mass-energy, and hence, denial
of gravity itself!  In the Newtonian case, one can see this from the
static virial theorem:
\begin{equation}
\Omega + 3 \int_0^{{R_0}} p ~ 4 \pi R^2~dR =0
\end{equation}
which shows that if $p\equiv 0$, one must have $\Omega \equiv 0$.
Recently, GR version of such virial theorem has been obtained for
the first time (Mitra 2006c) and it shows that, in GR too, {\em
denial of pressure  for a fluid means denial of self-gravity}.
Actually, whether in Newtonian or in Einstein or in any other
gravity, it must be so because of sheer thermodynamical reasons:

By thermodynamics, the equation of state (EOS) of the fluid under
consideration may be represented as $p \propto \rho^\Gamma$ and
conversely $\rho \propto p^{1/\Gamma}$. Then clearly a  $p \to 0$
situation implies $\rho \to 0$. And if $\rho =0$, we would have
$M=0$ and so would be the self-gravity; $\Omega =0$. Thus from
physics point of view, if the only {\em exact} solution of collapse
of a uniform dust is really to be considered as a mathematical
physics problem, the mass of the BH formed therein is $M=0$. Since
the proper comoving proper time of its formation is $\tau \propto
M^{-1/2}$, clearly, the actual value of $\tau_{EH} =
\tau_{singularity} \to \infty$ in this problem. This understanding
has several ramifications. When one erroneously assumes $M >0$ for
this idealized collapse, one obtains the traditional BH paradigm
where the finite mass BH is formed in prompt free fall time scale
while the time scale of its formation as recorded by an astronomer
is infinite ($t_{EH} =\infty$). In other words although an
astronomer would never see a BH, it is supposed to be formed only
w.r.t. a local observer outside the realm of observed physical
universe. This is at variance with the principles of all theories of
relativity, Galilean, Special or General. The spirit of relativity
is that ``rulers'' and ``clocks'' could indeed be different for
different observers and  so could be the measured numerical values
of relevant physical quantities. However, if the physical phenomenon
is observable to one observer, it must be observable to all, if it
is not observable to a given observer, it must not be observable to
any other observer either. In other words, all observers are on
equal fundamental footing and there is no favoured or disfavoured
observer. Physically, it means that all physically valid coordinate
systems must be connected by {\em non singular} transformations.
This fundamental spirit of relativity and physical rationality gets
violated in the traditional BH paradigm. This basic inconsistency
associated with the supposed traditional BH formation picture gets
resolved by realizing that actually $M_{BH}=0$ and $\tau_{EH} =
\tau_{singularity}= t_{EH} =t_{singularity}= \infty $. In other
words, neither the comoving observer nor the astronomer sees either
the EH or the central singularity formation as they indeed happen
asymptotically.

 In GR collapse equations, physics appears when one considers (i)
an EOS of the collapsing fluid, (ii) the evolution of the EOS at
arbitrary high pressure, temperature or other relevant parameters,
(iii) generation of heat/radiation within the body due to
dissipation and (iv) propagation of the outward radiation and its
effect on the collapsing fluid in a self-consistent manner. Without
such physical aspects, the GR collapse problem becomes a mere
exercise in applied/numerical mathematics something like the
dissection of a corpse without any flow in the veins and a throbbing
heart. Yet starting from OS, even now, many authors avoid all such
intractable physical aspects by setting $p \equiv U \equiv 0$ in the
problem. Technically, this is known as ``dust approximation''.
Formally, one can also assume the ``dust'' to be inhomogeneous
instead of the homogeneous case considered by  OS . But once one
assumes inhomogeneity, infinite forms of inhomogeneity (even though
density would be assumed to decrease with $R$) can be assumed. Then
complexity of the GR collapse equations offer infinite scope for
various additional applied mathematics exercises. In particular, in
some cases, it is claimed that, instead of a BH with an EH, there
may be ``Naked Singularity'' where the central singularity forms
before the central region gets trapped and hence, the  central
singularity may be visible either momentarily or permanently (Joshi
2004). If truly so, this phenomenon would be in the arena of
observable astrophysics. However such discussions never consider
physical questions such as (i) the value of $z$ at the moment of
formation/eventual of the supposed Naked Singularity, (ii) 3-speed
and acceleration of the fluid  and (iii) Mass of the Singularity
though in some cases it is believed that it would be have $M=0$.
Even if one would assume, $M=0$ for such Naked Singularities, one
wonders why they must not hurtle to $ M = -\infty$, in the absence
of an EH. In other words, why formation of a supposed global naked
singularity of any mass must not be followed by an infinitely strong
and {\em  visible} energy  outburst. Such physical questions are
rarely even posed, let alone answered, in the context GR cum applied
mathematics research (Joshi 2004).

We may point out that such questions too get resolved by realizing
that $p\equiv 0$ actually implies $\rho \equiv 0$ so that all dust
collapse problems must correspond to {\em uniform} density $\rho =0$
OS problem which uniquely and mathematically, results in a
mathematical BH of $M=0$ (physically, a $M=0$ BH is never formed as
the collapse process becomes eternal with $\tau =\infty$). Since the
sound speed within a fluid is $c_s = \sqrt{dp/d\rho}$, in order that
$c_s$ is finite, it is necessary that $d \rho =0$, if $dp =0$. Hence
from, this consideration too, there cannot be any true
``inhomogeneous dust'' and no Naked Singularity in dust collapse.

To settle this question permanently, let us recall the GR Poission
 equation for a static fluid (Ehlers, Ozsavath and Schucking 2006):
\begin{equation}
 \nabla^2 \sqrt{g_{00}} = 4 \pi G \sqrt{g_{00}} ~ (\rho + 3p)
 \end{equation}
%\end{document}
 where $g_{00}$ is the time-time component of the metric
coefficient defining both gravitational redshift and relativistic
potential. If the fluid would tend to become pressureless dust, it
would tend to undergo free fall where it would be possible to set
$g_{00} = uniform$ (actually unity) over the entire fluid. Thus, as
$p \to 0$, $\nabla^2 \sqrt{g_{00}} \to 0$ too.

%\end{document}

 Then it follows from the foregoing equation that $
\rho \to 0 ~as~ p \to 0 $. Thus all dust collapse, in reality,
correspond to the OS collapse which in turn corresponds to the
formation of a $M=0$ BH.

\subsection{Physical Interpretation of Adiabatic Collapse}
Many authors attempt to inject partial physics into the GR collapse
problem by considering the fluid to possess a pressure. However,
tracking of the actual evolution of the  EOS, particularly at
arbitrary high density and temperature is impossible both in
principle and practice. Thus usually some mathematically amenable
fixed EOS or modestly changing EOS is considered for
analytical/numerical studies. Further, in many cases, propagation of
radiation through the fluid is not considered at all by setting the
heat flow flux $q\equiv 0$. Yet, one can have infinite variations of
input parameters here to obtain various results such as finite mass
BHs or Naked Singularities as the end product. But do such results
have any physical validity in the absence of consideration of
radiation generation and its transport?

 The recently developed GR virial theorem (Mitra 2006c) shows that, for slow
 gravitational contraction
 \begin{equation}
 dQ = {(3\Gamma -4) \over 3(\Gamma -1)} ~ |d \Omega|
 \end{equation}
 where, $dQ$ is the amount of heat radiated by the collapsing body
 and
 $|d \Omega|$ is the magnitude of the change of the
 self-gravitational energy of the fluid.

 For strictly adiabatic radiationless collapse, $dQ \equiv 0$, and
 thus $\Omega$ must remain fixed during such contraction. But in
 spherical geometry, $\Omega$ cannot remain fixed during collapse
 unless it is pegged fixed at $\Omega \equiv 0$. Thus despite
 detail applied mathematics alongwith some semblance of physical
 considerations, a strict adiabatic collapse  implies $\Omega \equiv
 0$, a condition which is satisfied only for $p = M =0$. In such
 a
 case,
 despite the mathematical consideration of a pressure, actually,
 $\rho=0$.
  Hence despite many apparent richness of pressure aided
 adiabatic collapse, physics wise it is equivalent to the dust collapse
 if one must pretend a ``collapse'' in such a case.
 Therefore, despite the appearances of either finite mass BHs or Naked
 Singularities through applied mathematics exercises, adiabatic
 collapse should uniquely result in a zero mass BH. Technically a zero mass BHs
 is a borderline case between a BH and a Naked Singularity.
 Physically it is never formed since $\tau =\infty$. Even more
 physically, there is no strict adiabatic collapse and all
related applied mathematical results are just a chimera devoid of
physics.

If a given collapse  problem would claim to find emission of
radiation, then, by definition, one must use the formalism of
radiative collapse which necessarily involves heat flux $q$. The
local value of  luminosity would be $L = 4 \pi R^2 q$. Conversely,
any (adiabatic) collapse study which sets $q=0$ beforehand, must not
find any finite value of $L$ (by definition). However, in a peculiar
case of mathematical treatment, by introducing negative pressure and
fudging of the boundary conditions, one might claim to predict
``strong burst of radiation'' (i.e., very large $L$) in a purely
adiabatic collapse with $q\equiv 0$ (Joshi \& Goswami 2005,
Bojowald, Goswami, Maartens and Singh 2005, Goswami, Joshi and Singh
2006)! Such examples show that a mere applied mathematics treatment
of the GR collapse equations can not only be physically vacuous but
could be completely misleading too.

\subsection{Physical Interpretation of Radiative Collapse}
The amount of energy radiated in a continued or any collapse, $Q$,
is to be determined from self-consistent solution of GR collapse
equations. However the  GR virial theorem could roughly indicate
this amount from overall  thermodynamical considerations. And when
one is purporting to probe a likely singular state with infinite
density, pressure and temperature, one must not make crucial
simplifying assumptions about the value of $Q$. However, many
radiative studies of GR collapse, either for numerical or analytical
simplicity, implicitly or explicitly assume beforehand $ Q \ll M_0
c^2$. From the point of view of final stages of collapse, such an
assumption effectively reduces the problem to the adiabatic case,
and in turn, to the dust collapse case. To confirm this we recall a
recent generic result on contraction of self-gravitating
configurations (Mitra 2006a):
\begin{equation}
{\rho_r\over \rho_0} \sim z \gg 1; ~when~ \qquad z\gg 1
\end{equation}
Note that the body becomes a radiation dominated early Universe like
fireball {\em even before any EH would form}: $\rho_r \gg \rho_0$
for $z\gg 1$. The heat flux $q \propto \rho_r$ accordingly becomes
extremely large and the associated outward radiation force on the
plasma halts the collapse to form an ECO. Usually numerical
radiative collapse studies, on the contrary, assume $\rho_r \ll
\rho_0$. In view of the foregoing equation, this latter assumption
can be valid only in the regime of $z \ll 1$. As far as the regime
$z \gg 1$ is concerned, at best this assumption would imply $\rho_0
=\rho_r =0$. Thus true physical interpretation of such simplified
continued radiative collapse studies is that they correspond to
$\rho =0$ and a BH of $M=0$ though, by means of mere applied
mathematics, bereft of consistent physics and thermodynamics, they
may talk of generating finite mass BHs or Naked Singularities.

Thus even if one would consider idealized fluids such as a ``Scalar
Field'', thermodynamics demands that there is dissipation and heat
transport, uninhibited by any simplifying favorable assumptions, if
the fluid must  collapse. And if such dissipative processes and
heat/radiation transport mechanisms would be denied there would be
no collapse at all though by means of applied mathematics/numerical
computation one may arrive at variety of ``results''.

\subsection{Physical Gravitational Collapse}
Gravitational collapse must be accompanied by emission of
radiation/heat flow from the collapsing fluid. The density of
radiation within the fluid gets enhanced by (a) matter-radiation
interaction (diffusion) by which $\rho_r$ increases from the value
one would obtain using free streaming assumption (Mitra 2006a). And
in the $z\gg 1$ regime $\rho_r$ also increases (b) due to trapping
of the radiation by self-gravitation of the fluid (Mitra \&
Glendenning 2006). The process (a) has recently been specifically
considered. In a very important paper Herrera \& Santos (2004) have
shown that the effect of outward flow of heat can stall the GR
continued collapse. This pioneering suggestion has been confirmed by
Herrera, Di Prisco and Barreto (2006) by means of a specific
numerical modeling. In another related work, Cuesta, Salim and
Santos (2005) have found that Newtonian supermassive stars undergo
collapse to form a hot quasistatic ECO rather than a static cold BH.
Such works however have not considered the generic mechanism (b) of
self-gravitational trapping of radiation/heat. On the other hand if
this mechanism would indeed be implemented in an appropriate
numerical scheme, it would be found that ECOs rather than BHs are
formed for arbitrary initial mass of the fluid provided it is dense
enough to undergo continued collapse to $z \to \infty$.

\subsection{Comments on Some Numerical Works}
Following the suggestion by one of the referees, we shall
specifically point out why some numerical works lead to conclusions
completely different from what has been found here.

$\bullet$ 1. Let us consider the paper ``Collapse of a rotating
supermassive star to a supermassive black hole:  fully relativistic
solution'' by Shapirao \& Shibata (2002). GR is meant to be a
physical theory where all forms of mass energy couple with one
another and one may have the exact realization of the $E = Mc^2$
formula whereby entire passive initial gravitational mass  may
actually be transformed into pure energy (radiation). Normally the
nomenclature ``fully relativistic'' would mean that the concerned
study has incorporated all physical aspects of of the problem
without making any simplifying crucial assumption in the framework
of strongest possible gravity. But the actual reality is far from
this. Any number of crucial assumptions and simplifications are
often made in such studies and the nomenclature ``fully
relativistic'' is used from purely applied mathematic point of view,
i.e., the calculations are no  ``Post Newtonian'' ones. From this
this definition of ``fully relativistic'', the Oppenheimer Snyder
(1939) study which suppressed all physics by assuming $p =U \equiv
0$  even at the singularity is also ``fully relativistic''
calculation. Ironically, despite this, it is indeed the only {\em
exact} ``fully relativistic'' calculation because here
$p=\rho=q=L=M=M_0=0$.

Referring back to this work by Shapiro \& Shibata (2002), it does
not consider any radiation transport at all, $q \equiv 0$. Thus, as
discussed before, from thermodynamics point of view, eventually in
the regime of $z\gg 1$, it is just the Oppenheimer Snyder (1939)
collapse.

$\bullet$ 2. ``Collapse of a magnetized star to a black hole'' by
Baumgarte and Shapiro (2003). While this paper considers magnetic
field in the fluid, it does not consider any radiation transport.
Thus it has no relevance for final stages of physical continued
collapse. Further, Eqs.(1-3) used by this paper are the equations
for {\em free fall} which are strictly valid when magnetic field $B
=0$, pressure $p=0$, heat flux $q =0$! One definitely cannot expect
formation of hot ECO in such a study.

 $\bullet$ 3. ``Collapse of uniformly rotating stars to
black holes and the formation of disks'' by Shapiro (2004). This is
a Newtonian cum Post Newtonian study and does not consider any
radiation transport at all. Thus, in reality, it too has got no
relevance for final stages of  physical radiative continued
collapse.
\section{More Fundamental Reasons}
There are much more fundamental reasons for the GR  continued
contraction time scale to be $t=\infty$ than what has been
elaborated above.
\subsection{Speed at the Event Horizon}
STR is founded on the principle ``nothing can move faster than
light'' (actually faster than a certain limiting speed). GTR asserts
that this principle is valid even in curved spacetime, in the
presence of gravity, for an observer with arbitrary acceleration. It
is known for long that if a test particle would approach the EH, its
3-speed would approach the speed of light, $v \to c$ and in fact, if
one uses the so-called Schwarzschild coordinates, one has $v=c$ on
the EH (Mitra 2000, 2002a,b). If the BH would have, $M >0$, $R_g
>0$,  one would, in this case, have $v=c$ at $R=R_g >0$. If so,
the speed of the particle inside the EH would exceed the speed of
light, $v
>c$. Many GR ``experts'' try to fudge this real fundamental problem
by insisting that, in suitable coordinates, one may have $v_{EH}
<c$. This is actually impossible if the physics would be treated
self-consistently because, STR velocity addition law ensures that
once $v \to c$ w.r.t. a certain observer, it must be so w.r.t. any
other observer. Thus, at best one may assume that $v \to c$
asymptotically without ever exceeding it. This would be possible
only if $M=0$ and proper time to approach the EH, $\tau \to \infty$.
(Mitra 2000, Mitra 2002a,b). In technical parlance, existence of a
finite mass BH would violate the condition that the worldline of the
infalling particle must be timelike (Mitra 2000, 2000a,b, Leiter \&
Robertson 2003).

 It may be recalled that astrophysicists claiming to study the
 problem of accretion around supposed finite mass BHs indeed
(correctly)  insist that one must have $v=c$ at the EH (Chakrabarti
1996, 2001)
 unmindful of the fact that this is not allowed in finite mass BH paradigm
  and many
 BH ``experts'' struggle hard to suppress this fact! In an ironical
 situation, such astrophysical works which find $v=c$ at the EH, in
 blatant violation of the BH paradigm  claim to have found ``evidences for
 astrophysical BHs''! But no ``BH expert'' would ever point out that such works
 are in violation of the BH paradigm  and on the other hand  claim
 that the BH paradigm has been confirmed by astrophysical studies!
 \subsection {Acceleration at the Event Horizon}
 It is also known for long that radial component of the 3-acceleration $a^R$ of the test particle
 blows up at the EH. This profound physical result, inconvenient for the BH paradigm, has
 traditionally been brushed aside by mentioning that suitable
 coordinate transformations may remove this singular acceleration by
 ignoring the simple STR result that if acceleration is infinite in
 one frame, it is so in any other frame. Later it was pointed out
 that one can construct an acceleration {\em scalar} which too
 behaves in exactly the same singular way. This unequivocally showed
 the physically singular nature of the EH, which implies that the EH
 itself is the central singularity (i.e., $M=0$) (Mitra 2002a,b).
 However ``BH experts'' found such conclusions inconvenient and
 pretended to be unaware of it. Recently MacCallum (2006) has admitted that
 the acceleration scalar indeed blows up not only at the horizon of
 only Schwarzschild BH but for all BHs, for instance, spinning Kerr
 BH. However, in order to still uphold the BH paradigm, he has
 suggested new {\em mathematical criterion} for the definition of
 spacetime singularities completely ignoring that any such new rule
 would not change the basic fact that the physically measurable
 acceleration would blow up at the horizon in a {\em coordinate
 independent} manner. Recall, the original claim for supposed spacetime
 regularity at the EH was that ``no singular unusual physics happens
 there''. Thus clearly this attempt by MacCallum to uphold the BH
 paradigm is unjustified and inconsistent from the point of view of
 the accepted notion about a ``regular event horizon''. At this rate, one can claim that attainment of speed of
 light by a material particle is no violation of GR. In reality,
 atleast for a material test particle attainment of infinite
 (scalar)
 acceleration  and infinite Lorentz factor ($v=c$, $\gamma=\infty$) are closely
 related phenomenon.
\subsection{Non occurrence of Trapped Surfaces}
The singularity theorems buttressing the BH and Naked Singularity
paradigms are based on the assumption that trapped surfaces are
formed in continued spherical GR collapse. But it was shown very
transparently that trapped surfaces do not form in GR collapse
(Mitra 2004a, 2006f):
\begin{equation}
{2G M(r)\over R} \le 1
\end{equation}
where $r$ is the comoving radial coordinate. In contrast, formation
of trapped surfaces demand $ 2GM(r)/R
>1$ and the equality sign denotes formation of an ``apparent horizon''.
Eq.(58) shows that under the condition of positivity of mass, one
must have $M \to 0$  as the singularity would be approached $R \to
0$. If one would have $2GM/R <1$ at the singular state, there would
be no horizon, and $M$ should wander towards the negative branch.
Thus one must approach the equality limit of a zero mass BH : $2 GM/
R \to 1$ as $R \to 0$. However since the worldlines of the fluid
must always be timelike, this limit which corresponds to a {\em
null} condition must not ever be attained. In other words, one must
have $\tau \to \infty$ as $R \to 0$. We saw this categorically for
the dust collapse.

The physical reason for non-occurrence of trapped surfaces is again
``nothing can move faster than light'' because it was shown that
occurrence of a trapped surface would mean the local 3-speed of the
collapsing fluid would exceed the speed of light $v> c$ (Mitra
2004a, 20056e). In fact, at the ``apparent horizon'', one should
find the acceleration of the fluid to blow up.  Therefore, an
apparent horizon cum EH may only asymptotically form as $R \to 0 $
and $M \to 0 $. Unfortunately the BH and singularity ``experts'',
having already made too much commitment and investment in the
BH/singularity paradigm, found both the exact derivation of the
theorem of non-occurrence of trapped surfaces and its physical
interpretation to be most inconvenient and chose to quietly ignore
them. Since an EH is never formed, there is no trapping of any
quantum information within any EH. Thus actually, there was never
any Hawking radiation or Quantum Information paradox. And obviously
there need not be any resolution of this paradox  either because, in
reality, it was never there  because BH mass $M\equiv 0$ (Mitra
2006f).
\subsection{Ultimate Result}
By using the basic rule of differential geometry and curvilinear
coordinate transformation that the {\em proper 4-volume} for a
Schwarzschild BH must remain same in all coordinates, it has been
directly shown that the Schwarzschild BHs have the unique mass
$M\equiv 0$ (Mitra 2005, 2006d,e,f). It has also been shown that,
the rotating Kerr BH has the same fate (Mitra 2004b,c). Thus one can
physically understand the result of McCllum (2006) that acceleration
scalar blows up at the horizon in the latter case too.

But independent of the above result,
 following the field theoretic analysis of Arnowitt, Deser and
 Misner (1962), one can find that $M=0$ for a neutral  BH.
 The vacuum Schwarzschild solution describes the spacetime strucure
 of a neutral {\em point mass}.
 And  the ``clothed mass'' of a sufficiently small static neutral
 sphere  of radius $R_0$ is:
 \begin{equation}
 M = G^{-1} \left( -R_0 + [R_0^2 + 2 M_b ~R_0~ G]^{1/2}\right)
 \end{equation}
 where $M_b$ is the ``bare mass''. When $R_0 >0$, of course, $M >0$.
 But as $R_0 \to 0$ to become a ``point mass'', $M \to 0$  due to negative self-gravity
 in exact accordance with our results.

 \section{Black Hole Electrodynamics?}
 A Schwarzschild BH is just a {\em point mass} and  its all vacuum
 outside it even if we momentarily admit of the (incorrect)
 possibility $M >0$. In pure classical vacuum, electromagnetic waves
 can of course propagate. In such a case, one obtains a ``vacuum
 impedance'' of $\mathbb{I}\sim 377$ Ohms. But this ``impedance'' is no
 measure of vacuum resistance  $\mathbb{R} \neq \mathbb{I}$  and the
 latter is
  always $\infty$ in the absence
 of any free charge. Thus while electromagnetic waves can propagate
 in a perfect classical vacuum, {\em no current can pass through}
 the  same. One may ask then how does there is an electric discharge
 between two parallel {\em conducting} plates if sufficiently high
 external electric field is applied between them.  This is so because
 a
 sufficiently strong electric field  would {\em pulls out free electrons,
 ions} from the {\em conducting plate}.
 It is these released free electrons which modify the vacuum and become
 the carrier of the current. However  not a single electron is added  to the discharge
 by the original vacuum.

  If on the other hand, one would replace the {\em conducting}
  electrodes by two {\em perfect insulator}, there will be no vacuum
  current howsoever strong the applied field may be because
  $\mathbb{R}= \infty$. Further, in the previous case, even when the plates are
  physical conductors with abundant source of free charges,
  there will be no current if {\em the intervening region would be
  trapped from which nothing can escape}. In such a case, there can
  be only inward flow of electrons inside the {\em trapped region}
  and there cannot be any outward flow of charge by the very
  definition of Event Horizon and {\em trapped surfaces}. This
  inward  discharge would continue until the entire source of free
  electrons gets depleted. Therefore there cannot be any steady or
  even momentary  current outflow from a BH.

  Again going back to the previous example, even if there would be
  no trapped region between the conducting electrodes and the free
  electrons of the {\em electrodes and not of the vacuum} would
  constitute a steady current, there would be no electromagnetic
  coupling of the intervening media (vacuum) with the electrodes.
  Thus there is no question of energy extraction from a neutral
  nonconducting ``point particle'' sitting at the centre of the
  vacuum. On the other hand, such a coupling would be established
  only when the intervening medium is indeed a medium with either an
  intrinsic  magnetic field or an induced magnetic field due to some
  unipolar induction mechanism. The latter would require that the
  intervening medium is a spinning conductor and then, indeed,
  there can be a real electromagnetic coupling between a conducting
  Neutron Star or any star with its accretion disk.

  Only quantum electrodynamical processes may lend a pure vacuum
  some electrodynamic properties as it happens for the Cashimir
  effect between two close by glass plates. And this happens without
  the aid of external electric or magnetic field. A horizon can only
  suppress all electro dynamical properties by screening the central
  matter/plasma.

  $\bullet$  However, the astrophysical BH paradigm was built by
  blurring the physics and GR by means of wishful
artifacts like
  imagining the effective existence of a physical conducting ``membrane'' having finite redshift
  in place of an
 EH, an imaginary 2-sphere without any physical surface or conducting
  properties and having infinite redshift. This was done
   by ``stretching'' the EH to suppress  the physically singular
  behavior of the EH (Thorne, Price \& Macdonald 1986).  In fact,  these authors admitted that

(i) ``the velocity with which these FFOs see the FIDOs ..becomes the
  velocity of light at the horizon'' (pp. 22)

  without realizing that this is a singular behavior.

  (ii) ``A FIFO at the horizon measures a divergently large
  gravitational attraction.$g$... but he normalizes $g$ to a
  per-unit-universal-time basis, he obtains a finite
  acceleration'' (pp. 97)

  This is nothing but distortion of physics by dividing  one $\infty$ by another $\infty$
  to pretend that physics is regular at the horizon.

(iii) ``The mental  deceit of stretching the horizon is made
mathematically viable, indeed very attractive, by the elegant set of
membrane-like boundary conditions to which it leads at the stretched
horizon'' (pp. 46)

 The phrase {\em mental deceit} says it all. In one of the crucial
foundational papers of this paradigm where it is claimed that a
spinning BH immersed in an external electromagnetic field would
develop ``eddy currents'', Damour (1978) admits that

``From a phenomenological point of view it is convenient to {\em
introduce} a surface charge density and a current on the horizon.
The heuristic justification'' is

``Therefore if we {\em wish} to keep the charge and current
conserved...''

Essentially, it is presumed beforehand that a BH, as the supposed
central engine of quasar must electromagnetically interact with
magnetic field of the accretion disk, the way a {\em truly
magnetized} pulsar would do. Having made this presumption, the rest
of the scholarly theory is built based on ``phenomenological'' and
``heuristic'' means. In the same vain, if one would demand that
``there must be a conserved current threading any perfectly
insulating sphere'', one would obtain  a ``surface current'', a
``surface charge'' and all other attributed properties of a neutral
BH. And one would eventually obtain the magnetic spin down
luminosity of any spinning non-magnetized insulator sphere! Further,
subsequent researchers might think that the novel concepts like
``surface charge'' and ``surface current'' for a perfect insulator
were derived from the first principles.

Later since most authors  would require a framework to explain
astrophysical phenomenon, they would  use such ``results''  ever
delving deep into their roots. In the absence of alternative
physical theories, the authors will not have any option either. But
now that we know that $M=0$ for Schwarzschild BHs, it may be
realized that this approach was incorrect even if one would naively
accept that $\mathbb{R} = 377$ Ohm when for a BH or any pure vacuum,
actually, $\mathbb{R}  = \infty$.

In contrast a MECO as a hot ball of magnetized plasma and without a
horizon would be the ideal central engine for most of the high
energy processes in astrophysics.

\section{Conclusions}
Since the basic cause of energy liberation in astrophysics is
self-gravitation, a star or any other self-gravitating body, upon
exhaustion of thermonuclear or any other specific external source of
energy, cannot be completely devoid of supply of energy. Further, as
first correctly shown by Eddington, in the ultimate analysis, the
reservoir of energy of a star or any self-gravitating object is its
total mass energy. When one applies only special theory of
relativity, it appears that the minimum time scale for depletion of
this reservoir is a finite number determined only by fundamental
constants. This time scale may aptly be called ``Einstein-
Eddington'' time scale. However once we consider GR, in principle,
even the minimum value of Einstein - Eddington time scale could be
infinite. We found that it is this Einstein -Eddington time scale
rather than the pressureless free fall time scale (whose comoving
value remains same in both Newtonian and GR cases) is a measure of
the contraction time scale of self-gravitating bodies. The extremely
large value of the radiative time scale (see Eq.[37]) corresponding
to a measureable luminosity has got important physical significance.

 Recently, it has been shown that the during the final stages of BH formation ($z \gg 1$),
 the collapsing object would be dominated by radiation energy rather than by
 rest mass energy density (Mitra 2006a). For such a self-gravitating ball of radiation,
 the luminosity would indeed be maximal, i.e., $L \to L_{ed}$ or $\alpha \approx 1$.
 Consequently, the observed duration of such final stages would indeed be determined
 by Einstein -Eddington time scale as obtained in this paper
 (Mitra 2006b, Leiter \& Robertson 2003, Mitra \& Glendenning 2006).
 The fundamental reason that both the observed Einstein Eddington
 time scale as well as the comoving proper time scale for formation
 of the eventual zero mass BH is infinite is that trapped surfaces
 are not allowed in collapse of isolated bodies and, atleast for
 isolated bodies, GR is a singularity free theory even at the classical non -quantum level
  as cherished by its founder Einstein. The physical reason for the
  non-occurrence of both trapped surfaces and finite mass BHs is the
  same relativistic adage ``nothing can move faster than light''.
  One cannot but recall at this juncture that Einstein (1939) too attempted
  to disprove the existence of BHs by using the same adage. However
  he failed to properly recognize that the vacuum Schwarzschild
  (actually Hilbert) solution indeed suggests formation of unique
  zero mass BHs. On the other hand, since he tried to be aloof towards both the
  implications of this important solution and also the exact OS solution,
  his attempted disapproval
  of BH looked inconsistent and suspicious. Many BH/singularity ``experts''  of present epoch
  however take unkind
  advantage of this situation and often try to portray Einstein as
  a scientist who lacked sufficient appreciation of GR (Baez \& Hillman 2000):

``{\sl In 1939, Einstein publishes a paper which presents a rather
desperate (and entirely incorrect) argument that no body could
collapse past its Schwarzschild radius. The nature of the conceptual
errors in this paper show that Einstein still did not understand
either the distinction between a coordinate singularity (the
boundary of a coordinate chart)and a geometric singularity, nor the
distinction between local and global structure. (Indeed, there is no
evidence that Einstein ever understood correctly the geometry of all
exact solutions to his field equations).}''

   However the present paper and other relevant papers have shown that
   Einstein was actually correct contrary to the presently accepted view. And we
  are certain that with the development of astronomical observational
  techniques, probably in next 10-20 years, it would be recognized
  by all
  that

  $\bullet$ Einstein's physical intuition about non-existence
  of (finite mass) BHs was correct though he could not see (zero mass)
  BHs as the asymptotical solutions of physical continued
  gravitational collapse of a chargeless fluid. However, with
  regard, to a point particle possessing a charge, Einstein \& Rosen
  (1935)
    wrote that

 `` It also turns out that for the removal of the singularity it
  is not necessary to take the ponderable mass $m$ positive. In
  fact, as we shall show immediately, there exists a solution free
  from singularities for which the {\sl mass constant $m$ vanishes}.
  Because we believe that these {\em massless solutions are the
  physically important ones} we will consider here the case $m=0$''
  (emphasis is due to  the author).

  $\bullet$ Most of the present day BH/Singularity ``experts'' and
  many of the GR experts having, in some cases, more mathematical/numerical skill than
  Einstein were actually experts on either Differential Geometry, or
  Applied Mathematics relevant for GR studies or Numerical
  Computations riding on GR and not necessarily  on the
  intricate and subtle physics lying at the throbbing heart of GR.

  Also it would be recognized that such experts sustained the BH paradigm
  by ignoring/avoiding
  consideration of physically measurable 3-speed and acceleration at
  EH or apparent horizons  and by
  blurring the physics/thermodynamics/radiation transport aspects
  in the gravitational collapse problem.

  $\bullet$ Despite Eddington's unjustified public denouncement of
  Chandrasekhar's correct result on upper limit of {\em cold}
  self-gravitating objects, Eddington's physical intuition and
  insight were far superior to that of Chandrasekhar; he was the first to
  correctly visualize the unphysical Nature of (finite mass) BHs and
  insisted that

``I think there should be a law of nature to prevent a star from
behaving in this absurd way''

And as emphasized by Mitra (2006b) and Mitra \& Glendenning (2006),
this ``law of Nature'' is nothing but the  bending of radiation due
to strong self-gravity and consequent attainment of a critical
Eddington luminosity.

  Of course, at that time, Eddington
   too failed to recognize that the gravitational contraction process must
  be radiative and a  BH (with $M=0$) should indeed be the asymptotical solution of the
  continued collapse process. It would be recognized much later that
  Chandrasekhar's result about upper limit of {\em cold} objects was
  almost universally misinterpreted, most notably by Chandrasekhar himself,
   as an upper limit on mass of all
  compact objects, {\em hot} or cold. Thus it would be recognized
  that Chandrasekhar's discovery had a profound retrograde effect on the
  development of the
  physical theory of continued gravitational collapse and
  relativistic astrophysics in general. Probably this
  misinterpretation alongwith the misinterpretation that the OS
  collapse was physical and suggested formation of finite mass
  BHs (when in reality, there is no collapse without finite pressure
  and heat flux, or, mathematically, $M=0$ in such a case), put the
  clock back by 60 years as far as the question of the final state of
  continued collapse is concerned.

  It may be also recalled that the original idea of ``Relativistic
  Degeneracy'' was due to Anderson (1929); the original (crude)
  calculations about the upper mass limit of a (cold) White Dwarf
  was due to Stoner (1930); and the basic idea that White Dwarfs are
  supported by (cold) quantum degeneracy pressure was due to Fowler
  (1926). And Chandrasekhar jelled together such ideas with considerable
  mathematical rigor within the framework of Newtonian gravity.

  Finally, if a body is not undergoing continued collapse its
  lifetime against collapse is infinite. And if the body of
  arbitrary mass would be undergoing continued collapse because of
  sufficient density, its collapse time scale would be determined by
  Einstein -Eddington  time scale rather than by any free fall time
  scale. During the course of the collapse the object must become a
  hot throbbing dynamic ECO and continue to collapse eternally by avoiding the static cold BH
  stage as per the correct intuition of Einstein and Eddington.

Already there are tentative evidences that the stellar mass BH
candidates have strong intrinsic magnetic field in lieu of an EH
(Robertson \& Leiter  2002, 2003, 2004, 2006a) as was predicted
earlier (Mitra 1998, 2000a,b, 2002). There is also an {\em direct}
evidence that the central compact object of one of the most well
studied quasar Q0957+561 is a strongly magnetized ECO rather than a
BH (Schild, Leiter and Robertson 2006). Further, most of the
observations of the Sgr A*, the BH Candidate at the center of the
milkyway may  be explained by considering it a strongly magnetized
ECO instead of a BH (Robertson \& Leiter 2006b).

\subsection{Epilogue}
 General Relativistic  continued collapse is an eternal story of all objects trying to
``burn'' away their complete stock of mass  due to the grip of
self-gravity in ultimate realization of the $E= Mc^2$ formula. As
they ``burn mass'' they eventually become hot ECOs/MECOs having
neither any lower (except zero) nor any upper mass limit.

%\begin{acknowledgement}
{ \bf Acknowledgements}

  The author thanks Felix Aharonian for encouragement and MPI fur
  Kernphysik, Heidelberg, for kind invitation and hospitality. The author is
  also immensely thankful to the anonymous  Referee B for his/her
  open minded, unprejudiced and fair assessment of this manuscript
  which, at this unfortunate epoch of GR research, many others would find inconvenient and
  reject with mere pretexts. His/her suggestions led to considerable
  improvement in the overall clarity and strength of this
  manuscript.

%\end{acknowledgement}
 %\end{document}


\begin{thebibliography}{99}
\bibitem{} Anderson, W., Zs.f. Phys., 54, 433 (1929)
\bibitem{} Arnowitt, R., Deser, S. \& Misner, C.W., 1962, ``The
Dynamics of General Relativity'' in {\it Gravitation: An
Introduction to Current Research} (ed. L. Witten, Wiley, NY),
(gr-qc/0405109)
\bibitem{}  Baez, J. \& Hillman, C, 2000, see
http://math.ucr.edu/home/baez/RelWWW/ history.html
\bibitem{baum} Baumgarte, T.W. \& Shapiro, S.L., 2003, ApJ, 585, 930
\bibitem{bethe} Bethe, H. A., 1939, Phys. Rev.,  55, 434

\bibitem{boj} Bojowald, M., Goswami, R.,  Maartens, R. and Singh,
P., 2005, PRL, 95, 091302

\bibitem{} Bowers, R.L. \& Deeming, T., 1984, {\it Astrophysics I,
Stars}, (Jones and Barlett, Boston)
\bibitem{buch} Buchdahl, H.A., 1959,  Phys. Rev., 15, 1027
\bibitem{} Chakrabarti, S.K., 1996, MNRAS, 283, 325
\bibitem{} Chakrabarti, S.K., 2001 (Private Communication)
\bibitem{chandra} Chandrasekhar, S., 1967 {\it An Introduction to the Study of Stellar Structure}
(Dover, New York, 1967)


\bibitem{}   Cuesta, H.J.M.,  Salim, J.M. and  Santos, N.O., 2005, Paper
Presented in {\it 100 Years of Relativity}, Sao Paulo, Brazil,  see

http://www.biblioteca.cbpf.br/apub/nf/NF-2005.html

\bibitem{} Damour, T., 1978, PRD, 18(10), 3598

\bibitem{ed1} Eddington, A., 1920, Brit. Assoc. Repts., 45

%\bibitem[1988]{balluch} Balluch, M. 1988,
 %     A\&amp;A, 200, 58
%\bibitem[Kippen
\bibitem{ed2} Eddington, A., 1926,
{\it The Internal Constitution of Stars}, (Cambridge Univ. Press,
Cambridge, 1926)
\bibitem{} Ehlers, J., Ozsvath, I. \& Schuking, E.L., 2006, Am.J.
Phys., 74(7), 607
\bibitem{} Einstein, A. \& Rosen, N., 1935, Phys. Rev., 48, 73
\bibitem{} Einstein, A., 1939, Ann. Math., 40, 922
\bibitem{} Fowler, 1926, R.H., MNRAS, 87, 114
\bibitem{gos} Goswami, R., Joshi, P.S. and Singh, P., 2006, PRL, 96,
031302
\bibitem{} Herrera, L. \& Santos, N.O., 2004, PRD, 70, 084004
(gr-qc/0410014)

\bibitem{} Herrera, L., Di Prisco, A. and Barreto,
W., 2006, PRD, 73, 024008 (gr-qc/0512032)

\bibitem{joshi1} Joshi, P.S., 2004, preprint (gr-qc/0412082)
\bibitem{joshi2} Joshi, P.S. and Goswami, R., 2005, preprint
(gr-qc/0504019)
\bibitem{kw} Kippenhahn, A. \& Weigert,  A., 1990, {\it
Stellar Structure and Evolution}, (Springer, Berlin)
\bibitem{} Kumar, S.S., 1962, AJ,  67, 579
\bibitem{} Kumar, S.S., 1963a, ApJ, 137, 1121
\bibitem{} Kumar, S.S., 1963b, ApJ, 137, 1126
\bibitem{} Leiter, D.J., Mitra, A.   \& Robertson, S.L., 2001
(astro-ph/0111421) (the original version)
\bibitem{lr} Leiter, D.J.  \& Robertson, S.L.,
2003, Found. Phys. Lett., 16, 143
\bibitem{} McCallum, M.A.H., 2006, Preprint (gr-qc/0608033)
\bibitem{} Misner, C.W. et al., 1972, PRL, 28, 998

\bibitem{MTW}   Misner, C. W., Thorne, K.S.,  and Wheeler, J.A., 1972, {\it Gravitation}
(W.H. Freeman, New York, 1973)

\bibitem{} Mitra, A. 1998a, Preprint (astro-ph/9811402)
\bibitem{} Mitra, A., 1998b, Preprint (astro-ph/9803014)



\bibitem{mit2000} Mitra, A., 2000, Found. Phys. Lett.,  13, 543,
(astro-ph/9910408)
\bibitem{mit2002}  Mitra, A., 2002a, Found. Phys. Lett.,  15, 439,
(astro-ph/0207056)
\bibitem{} Mitra, A., 2002b, Bull. Astron. Soc. India, 30, 173
(astro-ph/0205261, Conf. Proc., Invited Talk)
\bibitem{mit2004} Mitra, A., 2004a,  Preprint,  (astro-ph/0408323)
\bibitem{} Mitra, A., 2004b, Preprint, (astro-ph/0407501)
\bibitem{} Mitra, A. 2004c, Preprint, (astro-ph/0409049)



\bibitem{} Mitra, A., 2005, Preprint (physics/0504076)

\bibitem{mit1} Mitra, A., 2006a, MNRAS Lett., 367, L66  (gr-qc/0601025)
\bibitem{mit2} Mitra, A., 2006b,  MNRAS, 369, 492, gr-qc/0603055
\bibitem{mit 2006c} Mitra, A., 2006c, Phys. Rev., D74, 0605066,
gr-qc/0605066


\bibitem{mit2005c} Mitra, A., 2006d, Advaces in Space Research (in
press),  (astro-ph/0510162)
\bibitem{} Mitra, A., 2006e, Oral Presentation, 11 th Marcel
Grossmann Meeting on General Relativity (Berlin)

\bibitem{} Mitra, A., 2006f, ``Black Holes or Eternally Collapsing
Objects: A Review of 90 Years of Misconceptions''  in {\it Focus on
Black Hole Research}, (ed. P.V. Kreitler, Nova Sc. Publishers, NY)
\bibitem{} Mitra, A., 2006g, Proc. 29th Int. Cos. Ray Conf., Vol. 13,
p.125 (physics/0506183)
\bibitem{} Mitra, A., 2006h, Proc. 29th Int. Cos. Ray Conf., Vol. 4,
p.187 (astro-ph/0507697)

\bibitem{mg} Mitra, A. \& Glendenning, N.K.,
2006, Oral Presentation, 11th Marcel Grossmann Meeting On General
Relativity (Berlin, 2006)





%ibitem[1938]{4} Bethe, H. A., 1939, Phys. Rev.,  55, 434
\bibitem{OS} Oppenheimer, J.R. \& Snyder, H., 1939, Phys. Rev., 56, 455
\bibitem{} Peters, P.C., 1970, PRD, 1(6), 1559

\bibitem{rl1} Robertson, S. \& Leiter, D., 2002, ApJ, 565, 447
\bibitem{rl2} Robertson, S. \& Leiter, D., 2003, ApJL, 596, L203
\bibitem{rl3} Robertson, S. \& Leiter, D., 2004, MNRAS, 350, 1391
\bibitem{rl2} Robertson, S. \& Leiter, D., 2006a, ``The Magnetospheric
Eternally Collapsing Object (MECO) Model of Galactic Black Hole
Candidates and Active Galactic Nuclei'' in New Directions in Black
Hole Research, ed. P.V. Kreitler (Nova Sc. Publishers, NY, 2005)
\bibitem{} Robertson, S. \& Leiter, D., 2006b, astro-ph/0603746

\bibitem{} Schild, R.E., Leiter, D.J., and Robertson, S.L., 2006,
AJ, 132, 420, (astro-ph/0505518)
\bibitem{shapiro} Shapiro, S.L., 2004, ApJ, 610, 913
\bibitem{shapiroshibata} Shapiro, S.L. \& Shibata, M., 2002, ApJ, 577,
904
\bibitem{} Stoner, E.C., 1930, Phil. Mag., 9, 944

\bibitem{} Thorne, K.S., Price, R.H., and, Macdonald, D.A., 1986,
{\it Black Holes: The Membrane Paradigm}, (Yale University Press,
New Haven)
\bibitem{vaidya}  Vaidya, P.C., 1951, Proc. Ind. Acad. Sc., A33, 264
\end{thebibliography}
\end{document}